\input psfig.sty
\documentclass[letter]{article}
\usepackage{graphicx,rotating,color}
\usepackage[draft]{kmn}
\usepackage{mnras_cite}

\newcommand{\til}{$\sim$}
\newcommand{\msun}{\thinspace\hbox{$M_{\odot}$}}

\newcommand{\pcmsq}{\thinspace\hbox{cm}^{-2}}
\newcommand{\ergsqcmsec}{\thinspace\hbox{$\hbox{erg}\thinspace\hbox{cm}^{-2}
                \thinspace\hbox{s}^{-1}$}}
\newcommand{\ergsqcm}{\thinspace\hbox{$\hbox{erg}\thinspace\hbox{cm}^{-2}$}}
\newcommand{\ergsec}{\thinspace\hbox{$\hbox{erg}\thinspace\hbox{s}^{-1}$}}
\def\degree{\nobreak\ifmmode{^\circ}\else{$^\circ$}\fi}
\def\spose#1{\hbox to 0pt{#1\hss}}
\def\simlt{\mathrel{\spose{\lower 3pt\hbox{$\mathchar"218$}}
     \raise 2.0pt\hbox{$\mathchar"13C$}}}
\def\simgt{\mathrel{\spose{\lower 3pt\hbox{$\mathchar"218$}}
     \raise 2.0pt\hbox{$\mathchar"13E$}}}
\def\targ{SAX J1808.4-3658}

\title[The optical counterpart to \targ: observations in quiescence]
{The optical counterpart to \targ: observations in quiescence}
\author[L. Homer, P.A. Charles, D. Chakrabarty and L. van Zyl]
{L. Homer$^{1,2}$, P.A. Charles$^3$, D. Chakrabarty$^4$ and L. van Zyl$^1$\\
$^1$ Department of Astrophysics, Nuclear \& Astrophysics Laboratory,
Keble
Road, Oxford OX1 3RH \\
$^2$Astronomy Department, University of Washington, Box 351580, Seattle WA 98195, USA\\
$^3$Department of Physics \& Astronomy, University of Southampton, 
 Southampton, SO17 1BJ, UK\\
$^4$Center for Space Research, Massachusetts Institute of Technology, Cambridge, MA 02139, USA\\
}
\begin{document}

\maketitle

\begin{abstract}
We report the first extensive set of optical photometric observations of the counterpart to \targ\ (V4580 Sagittarius) in quiescence.  The source was detected at
$V\sim21$, 5 magnitudes fainter than at the peak of its 1998 outburst.  However, a comparable \til6\% semi-amplitude 2hr modulation of its
flux is revealed.  This has the same phasing and approximately
sinusoidal modulation as seen during outburst, and with photometric minimum when the pulsar is behind
the companion.  The lack of a double-humped morphology rules out an ellipsoidal origin,
implying that the bulk of the optical flux does not arise from the companion.  Moreover, applying crude modelling to the disc and X-ray irradiated
face of the donor shows that the internal energy release of a remnant disc (with mass transfer driven by gravitational radiation) is sufficient
to explain most of the optical emission, and with the modulation due to the varying contribution of the heated star's face.  We note that this model is
also consistent with the much lower X-ray to optical flux ratio in quiescence versus outburst, and with the phasing of the optical modulation.
\end{abstract}


\section{Introduction}

The low-mass X-ray binary (LXMB) transient \targ\ was first detected
(in outburst) by the {\it Beppo}SAX Wide field Camera in 1996 September \cite{intZ98b},
following a non-detection in a preceding August observation.  Three type-I X-ray bursts were also observed demonstrating the
presence of a neutron star accretor. The third exhibited clear double-peaked morphology implying an Eddington-limited photospheric radius expansion event,
yielding a distance estimate of 2.5kpc \cite{intZ00}.

During another outburst in 1998 April/May, \targ\ was detected by {\it RXTE}, enabling the measurement of the fastest ever LMXB X-ray pulsation, with a
period of 2.49ms \cite{wijn98}.  \scite{chak98a} utilized precise timing of the millisecond pulses to determine many parameters of its
binary orbit: period, epoch of mean longitude, projected semi-major axis, eccentricity and pulsar mass function. Hence, they were immediately
able to place constraints on the nature of the system, in particular
the secondary's spectral type, suggesting a very low mass ($\simlt0.1$\msun),
but irradiation-bloated/ablated, dwarf star. Lastly, \pcite{roch98} were able to identify an optical counterpart consistent with the {\it
Beppo}SAX X-ray position, even within the smaller error circle yielded by the {\it RXTE} X-ray pulse timing.  We note that this counterpart has
recently been designated as the variable star
V4580 Sagittarius \cite{kaza00}.

Additional constraints on the secondary can be placed by observations of the optical counterpart. \scite{gile98} report on CCD photometry
performed with the Mt Canopus 1m telescope during decline from the
1998 outburst in April--June.  Their coverage consists mostly of single
pointings a few nights apart, in order to follow the long-term
lightcurve, which showed the source to fade from V=16.72 to around
18.5, 1 and 3 weeks after the start of the X-ray decline.  A later
observation on June 27 failed to detect the source, corresponding to
$V\simgt$20.5.  They also obtained short $V$-band sequences on three
nights, but only amounting to \til0.25, \til1.0 and \til1.2 cycles of
the 2 h period respectively. From these sparse data they concluded
that there was evidence for a 0.06 mag semi-amplitude modulation at
the 2 h X-ray period, with phasing such that the minimum occurs when
the companion lies between the observer and pulsar.

In order to place further limits on the secondary, we obtained the
first high time-resolution CCD photometry of the counterpart in quiescence, using the 1.9m telescope at the South
African Astronomical Observatory (SAAO), and it is these results that we present in this paper.

\section{Observations and data reduction}
\begin{figure}[htb!]
\begin{center}
\leavevmode
\resizebox*{.45\textwidth}{!}{\rotatebox{0}{\includegraphics{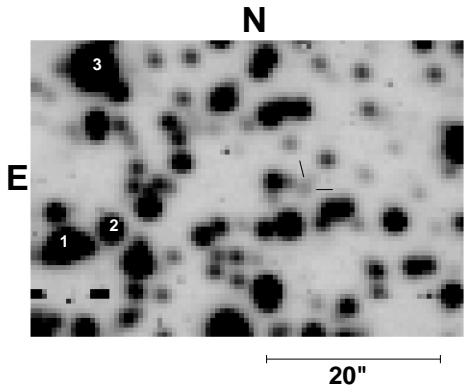}}}
\caption{Finding chart for V4580 Sgr (\targ) in quiescence. This is a co-add of
13 $B$-band images and totals 52 min of exposure.  Note the faintness
of the target and
and the closeness of neighbouring stars. The three local standard stars used for calibration are indicated. \label{fig:image}}
\end{center}
\end{figure}

Observations of a small ($50\times 33$ arcsecs) region surrounding the
optical
counterpart to \targ\ were made using the UCT-CCD fast photometer
(O'Donoghue 1995), at the Cassegrain focus of the 1.9m telescope at SAAO,
Sutherland on 1999 August 10 and 2000 July 3. 
  The UCT-CCD fast photometer is a   
Wright Camera 576$\times$420 coated GEC CCD which was used here
half-masked so as
to operate in frame transfer mode, allowing exposures of as short as 2s
with no dead-time. However, given the faintness of the quiescent counterpart we employed much longer integration times.  In 1999, we
concentrated on obtaining a high signal-to-noise lightcurve, and therefore we opted for a continuous time-series of 30s white light (no filter) exposures.
In 2000, we were again favoured with good seeing and stable photometric conditions
initially, so we attempted to obtain colour information on the binary modulation.  In fig.~\ref{fig:image} we show a deep (52 min exposure) $B$ band finding chart, constructed from the 13 shifted and co-added images from the interval of best
seeing during these observations.  This demonstrates the difficulties in obtaining good photometry of this faint target, as it is quite
severely crowded; only with good seeing, $\simlt1\arcsec$ (indeed truly excellent for this telescope) is there any possibility of resolving
the counterpart from its near neighbour.  However, the seeing did deteriorate and become more variable
towards the end of the run.  In this instance, we cycled through $BVR$ filters, with typical exposure times of 240, 180, 150s in each
respectively.  For the 2 h period, this still yields approximately 0.1 phase resolution.

We
performed data reduction using IRAF, including photometry with the 
implementation of DAOPHOT II (Stetson 1987).  Point spread function (PSF) fitting was always employed, as it is essential for obtaining the best possible
photometry, in such crowded conditions.  The details of this procedure
are given in Homer, Charles \& O'Donoghue (1998).  Lastly, in order to reduce systematic effects 
differential photometry was applied, wherein the 
magnitudes of a star are calculated relative to an ensemble of local standards (bright, non-variable stars). 

\section{Results}
\subsection{Colour photometry}
For the 1999 August 10 white light observations no strict calibration is possible.  However, if we assume that the unfiltered passband of the CCD corresponds roughly to $V$, we may use observations of the E-region standard E752 \cite{menz80} to estimate that the counterpart had $V\sim20$ on that date, consistent with the limit of \scite{gile98} given the uncertainties.

For the sets of colour images from 2000, we use the standard star E508 (chosen for its similar colours, and comparable airmass when observed),
and aperture photometry to derive the calibrated magnitudes of local standard stars within the small field of the CCD (see
fig.~\ref{fig:image}).  Finally, the magnitude of the target is calculated from its differential magnitude with respect to these local
standards, these measurements necessarily making use of PSF fitting.  We find $B=22.0\pm0.1$, $V=21.5\pm0.1$ and $R=20.9\pm0.1$ all uncorrected
for reddening. The magnitude errors quoted arise largely from counting statistics as the target is simply so faint, together with
uncertainties in the transformation between aperture and PSF
magnitudes.  Compared to its colours during outburst \cite{roch98} of $V=16.6\pm0.2$ and $R=16.1\pm0.2$, i.e. $(V-R)_{hi}=0.5\pm0.3$ cf. $(V-R)_{lo}=0.6\pm0.2$; apart from being 5 magnitudes fainter, they are consistent within the uncertainties, agreeing with the unchanging $V-I$ values found by \scite{gile98} during the decline itself.

Furthermore, in order to facilitate other comparisons we may attempt to correct for interstellar extinction towards \targ.  We utilize the
column density estimated from {\it RXTE} X-ray spectral fits by \scite{hein98} to set an upper
limit, and the galactic value in the direction of \targ\ for the lower
limit \cite{dick90}, to derive $6.3\geq N_H\geq1.3\times10^{21}\pcmsq$.  Applying the
relations $N_H=1.79\times10^{21}A_V$ \cite{pred95}, $E(B-V)=A_V/3.2$, $A_B=1.32A_V$, and $A_R=0.81A_V$, we may estimate corrected colours of
$-0.6\leq (B-V)_0 \leq 0.3$ and $-2.4\leq (V-R)_0 \leq 0.5$.  The corrected flux densities can also be obtained using the calibration for an
A0V star in 
\scite{cox00},
as may the absolute magnitudes assuming a $2.5\pm0.1$kpc distance
\cite{intZ00}. All these data are summarised in table~\ref{tab:optprop}.

\begin{table}
\caption{Quiescent optical properties of V4580 Sagittarius/\targ \label{tab:optprop}}
\begin{center}
\begin{tabular}{l l l l } \hline
 {Quantity} & Band & {Lower limit}&{Upper limit}\\\hline
Apparent mag.& $B_0$ & 17.7 & 21.0\\
(de-reddened)		& $V_0$ & 18.2 & 20.8\\
						& $R_0$ & 17.7 & 21.0\\
Absolute & $M_{{\rm B}_0}$& 5.7 &9.0\\
magnitude						& $M_{{\rm V}_0}$& 6.2 & 8.8\\
						& $M_{{\rm R}_0}$& 6.3 & 8.3\\
Flux density & $F_{B_0}$ & 2.5$\times10^{-17}$ & 5.3$\times10^{-16}$\\
(\ergsqcmsec&$F_{V_0}$ & 1.8$\times10^{-17}$ & 2.0$\times10^{-16}$\\
\hspace{1.5cm}\AA$^{-1}$)					&$F_{R_0}$ & 1.3$\times10^{-17}$ & 8.4$\times10^{-17}$\\
\hline
\end{tabular}
\end{center}
\end{table} 

Lastly, independent of both distance and reddening estimates we may compare the X-ray to optical flux ratios in outburst and quiescence.  We
use the outburst {\it RXTE}/PCA X-ray and contemporaneous optical observations on 1998 April 18 \cite{gilf98,roch98}, whilst we adopt an
intermediate value between the two quiescent X-ray measurements by {\it Beppo}SAX on 1999 March 17-19 \cite{stel00} and {\it ASCA} on 1999
September 17 \cite{dota00} together with our own optical data of 2000 July.  Hence, the quiescent value will be uncertain by a factor of a few,
given the known long-term X-ray flux variations.  The X-ray fluxes have been transformed to
the 2-10 keV range using W3PIMMS at HEASARC, and assuming a power-law spectrum with photon-index $\Gamma=2$.  We
calculate $F_X(2-10{\rm keV})/F_V=9.0\times10^{-10}/8.5\times10^{-13}=1100$ in outburst and
$1\times10^{-13}/9.3\times10^{-15}\sim10$ in quiescence where $F_V=990$\AA$\times f_V$, the flux density in $V$.  The ratio in outburst is comparable
to other outbursting transients and persistent LMXB, where the dominant optical emission is presumed to arise from X-ray reprocessing --
clearly this is not the case in quiescence.

\subsection{Orbital modulation}
\begin{figure*}[htb!]
\begin{center}
\leavevmode
\resizebox*{.9\textwidth}{!}{\rotatebox{-90}{\includegraphics{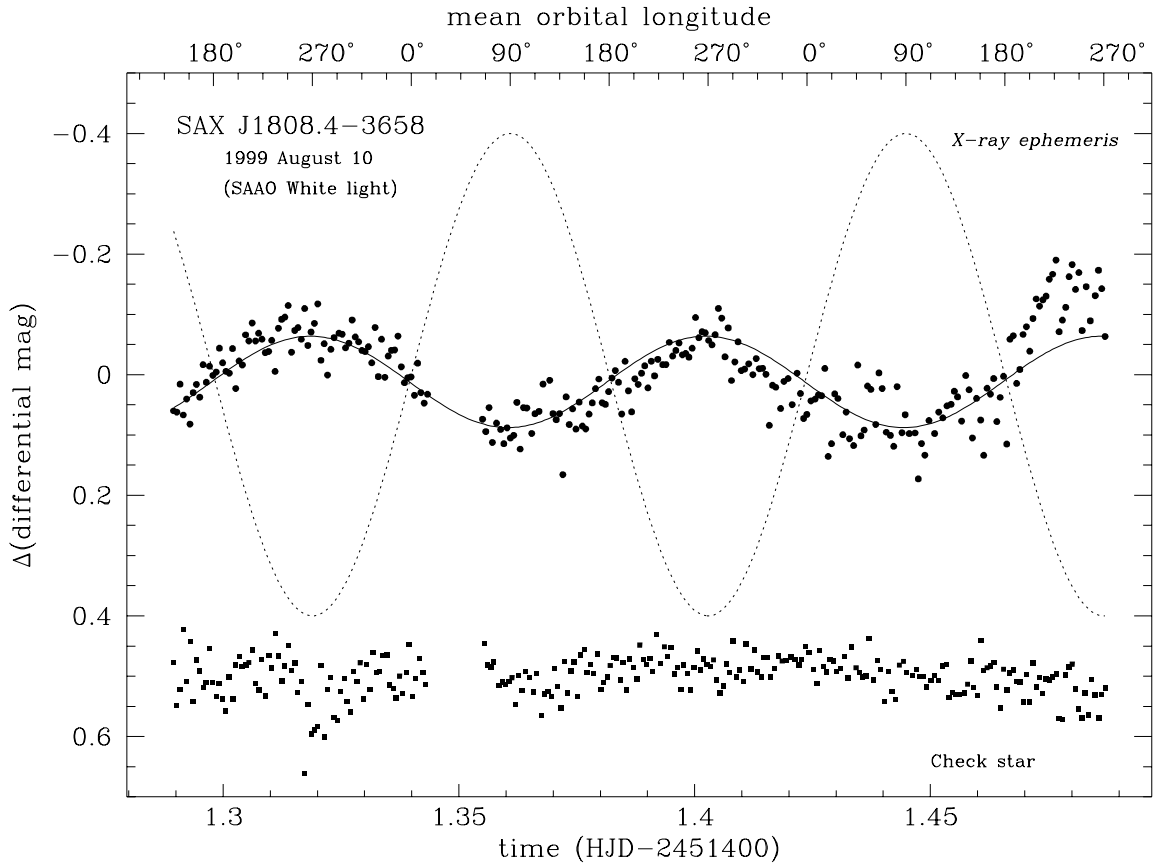}}}
\caption{Lightcurve of V4580 Sgr (\targ), taken in
white light (filled circles), with a star of comparable brightness plotted offset
below (squares).  The data has been binned up a factor of two to 60s time resolution for clarity. Over plotted is the sinusoidal fit to
the data (solid line), with only the period constrained to the orbital
value. The dotted line indicates the X-ray 
ephemeris, i.e. the mean orbital
longitude of the system as derived from the X-ray pulse arrival time delays (Chakrabarty \& Morgan 1998).  As
during the outburst observations of Giles, Hill \& Greenhill (1999), the time of photometric minimum is consistent with a mean orbital longitude of 90\degree,
that is when the pulsar lies behind the companion. \label{fig:WLlcs}}
\end{center}
\end{figure*}

\begin{figure}[htb!]
\begin{center}
\leavevmode
\resizebox*{.45\textwidth}{!}{\rotatebox{-90}{\includegraphics{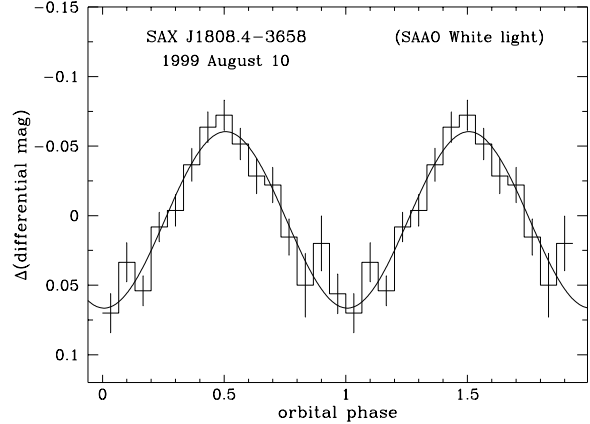}}}
\caption{Phase folded and binned lightcurve (white light) of V4580 Sgr (\targ), with a constrained sinusoidal fit over-plotted.
Two orbital cycles are shown for clarity. \label{fig:WLfds}}
\end{center}
\end{figure}

The lightcurve from 1999 August spans a full 2 cycles, with another half cycle, where the effects of airmass (and deteriorating seeing) become
a slight problem.  Clearly, this is insufficient to make any independent measurement of the period, but we can easily check for consistency with the pulse
timing result.  In figure~\ref{fig:WLlcs} we plot the linearly detrended lightcurve, along with a fitted sinusoid with $P_X=2.01$h and the projected X-ray
ephemeris.  Clearly, there is no doubt from our dataset that the quiescent counterpart exhibits a smooth approximately sinusoidal modulation
at the 2 h period, and the fit gives the semi-amplitude as $0.063\pm 0.003$ mag.  For this fit and later analysis we opted to omit the last half
cycle of data, where the source appears to reach its brightest level, as this may be due to the poorer seeing.  We note that the amplitude, like the colours, is consistent
with that of the high state source. The scatter seen in the similar lightcurves of \scite{gile98} could be due to an intrinsic property of the
high state source (it is known on occasion to exhibit complex X-ray flaring behaviour; \pcite{vdK00,wijn00}), which is likely to be present in the reprocessed optical
emission.  In order to study the morphology more closely, we have also folded and phase binned the data, which is presented in
figure~\ref{fig:WLfds}.  The pure sinusoid appears to be a very good fit, only at around 0.9 in phase does the data scatter, which corresponds to a region in phase for which only one cycle of data is present (owing to a
gap in the first cycle).  Quantitatively, we obtain $\chi_\nu=1.1$ (for 12 d.o.f) for the fit to the folded and binned data. 

\begin{figure}[htb!]
\hspace{-4.9cm}\resizebox*{.75\textwidth}{!}{\rotatebox{0}{\includegraphics{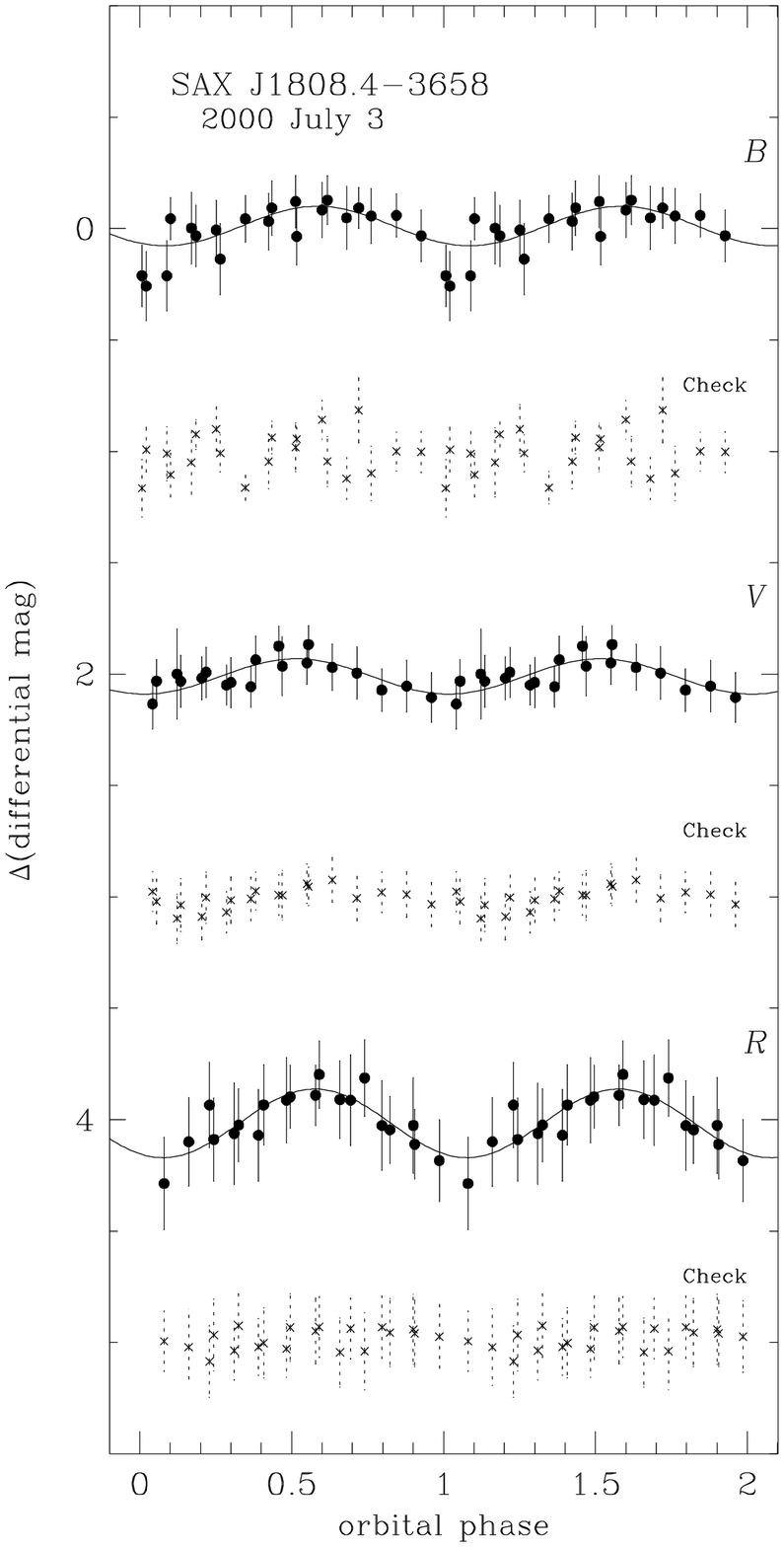}}}
\caption{Phase folded lightcurves ({\sl BVR}) of V4580 Sgr (\targ), with stars of comparable brightness plotted beneath
each. Fitted sinusoids with constrained period (=1.0) are overplotted on the data for V4580 Sgr. For
clarity two orbital cycles are shown and each curve has been offset by 1 mag. \label{fig:col_fds}}
\end{figure}

In July 2000 we attempted to investigate the colour dependence of the two hour modulation. However, the target turned out to be even fainter
than in August 1999, and hence the signal-to-noise of our data is marginal.  In fact, we found it necessary to firstly select the best seeing
images (based on check star behaviour), and then phase fold the data.  The data have also been detrended using a quadratic polynomial constrained to
approximate the airmass variations.  In figure~\ref{fig:col_fds} we
show the resulting $BVR$ lightcurves for V4580 Sgr and a star of comparable brightness.  Clearly, the phase folding reveals apparent
modulations in each band, although in the case of $V$ the check star also shows only slightly smaller amplitude variations.  Given the data quality we must be
cautious with regard to the significance of these modulations.  However, statistically we do find that a (period and phase constrained)
sinusoid fit is always an improvement over a constant value at $>99\%$ confidence level (from an  F-test). The fitted sinusoids also
provide marginal evidence for a trend towards larger amplitude at longer
wavelengths (see table~\ref{tab:amps}).  Even less certain is the possible ``eclipse'' feature present in the $B$ band data at phase\til0.0, which,
if excluded, does lend further support to a trend in the smooth modulation amplitude.  However, it is clear that further observations from a
larger telescope with better seeing are needed to confirm these tentative results.

\begin{table*}
\caption{Parameters of sinusoid fits to optical modulations in V4580 Sgr/\targ. \label{tab:amps}}
\begin{center}
\begin{tabular}{l l c c c} \hline
Lightcurve  & Phase of  & \multicolumn{2}{c}{Amplitude (mag)} & Sinusoid vs. constant fit\\
				& minimum light&Constrained: Period & Period and Phase& F-test confidence level\\
Whitelight  & $0.0084\pm0.0081$& $0.063\pm0.003$ &$0.063\pm0.003$ &100.00\%\\
$B$ band & $0.09\pm0.06$& $0.09\pm0.03$ &  $0.08\pm0.03$ & 99.18\% \\ 
$B$ -- ``eclipse''&$0.14\pm0.06$& $0.05\pm0.02$ &$0.03\pm0.02$ & 85.23\%\\
\hspace{.6cm} points&&\\
$V$ band&$0.02\pm0.04$& $0.08\pm0.02$ &$ 0.08\pm0.02$& 99.99\%\\
$R$ band&$0.08\pm0.02$& $0.15\pm0.02$ &$0.13\pm0.03$& 99.98\%\\
\hline
\end{tabular}
\end{center}
\end{table*} 

\section{Origins of the optical flux}
The detection of a plausibly phased two hour modulation confirms our identification of the correct counterpart, V4580 Sagittarius, in both sets of observations.
However, we have noted that its X-ray to optical luminosity ratio is very small, which implies that it is surprisingly optically bright,
especially given the expected very faint emission
from any secondary capable of fitting into the Roche lobe of the 2 hr
orbit (see below).
The possibility that the true quiescent amplitude is much larger than in outburst, allowing for another line-of-sight star,
responsible for some of the flux we 
have measured, can nevertheless be ruled out by the morphology, which is not flat-bottomed but sinusoidal (cf. the case of PSR 1957+20, \pcite{djor88}).
We shall now seek to apply some constraints as to the source of the quiescent optical emission, given the available information, although we are
limited by the large uncertainties in the absolute magnitude and colours of V4580 Sgr (due to the poorly constrained column).

Firstly, are we simply seeing the companion star and nothing else? \scite{chak98a} argue that the most probable companion is an extremely low
mass star ($\simlt0.1$\msun).  However, if this lies on the main sequence the expected $M_{\rm V}\sim14$ implies a brightness that is 5 magnitudes fainter than our
observed limit.  On the other hand, we might postulate a more massive
(0.5--0.8\msun) companion, such as a K1 to M0 main sequence dwarf, which would have the correct
magnitude.  But this is also not tenable. The counterpart is much too
blue for such a spectral type ($(B-V)_0\le0.3$ cf. a K1 star
value of 0.86).  The inferred inclination would be extremely low ($i\simlt6\degree$) and therefore improbable, since the {\it a priori} probability of seeing a
system with $i$ or less goes as $(1-\cos i)$.  Further compelling evidence is the lack of ellipsoidal variations.  If any modulation is to be seen from the light of the
companion star alone (as is the case for a number of other soft X-ray transients in quiescence), then its period should be half that of the
binary. Hence, we may conclude that no ``ordinary'' lower MS star provides all the optical emission.

\begin{figure}[htb!]
\begin{center}
\leavevmode
\resizebox*{.46\textwidth}{!}{\rotatebox{0}{\includegraphics{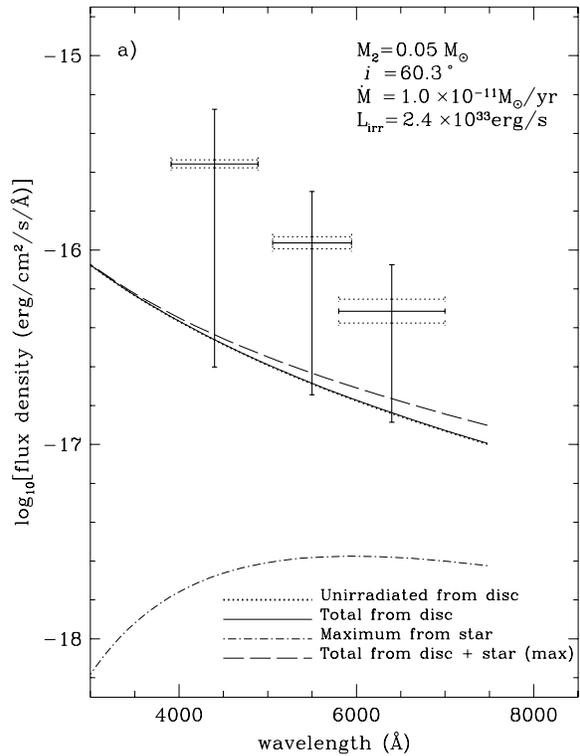}}}
\caption{Optical ($BVR$) photometry of V4580 Sgr (\targ) (points with errorbars), with model spectra from various (irradiated) disc/star emission
components (as labelled). Results for a secondary mass of 0.05\msun are shown,
which immediately determines the inclination to be $i=60.3\degree$.  The dashed errorbars indicate the maximum and minimum fluxes exhibited during the 2~hr modulation cycle, to be
compared with the variation between total from irradiated disc (solid
line) and total from irradiated disc + star (long dashed line). No attempt
has been made to fit the model to the data, instead reasonable values of $\dot{M}\sim10^{-11}$\msun~yr$^{-1}$ and $L_{irr}\sim10^{33}$\ergsec
have been chosen for illustrative purposes. \label{fig:spec.05}}
\end{center}
\end{figure}

The other possible source of optical emission is a combination of: (i) a remnant accretion disc, (ii) a hotter irradiated accretion disc,
(iii) the irradiated face of the companion star.  In order to make an approximate quantitative comparison with our data we have undertaken some
simplified modelling to obtain predicted broad band
spectra.  For the details of the modelling see \scite{chak98b}, \S4.3
(accretion disc), and the appendix (heated face of companion star).  The resulting spectra are
presented in figure~\ref{fig:spec.05}.  

Assuming that an accretion disc is still
present, how bright will it be solely due to internal viscous energy release?  Here
we make use of the results of the $\alpha$ disc prescription, using a nominal but reasonable value of $\alpha$=0.1, and a canonical 1.4$\msun$
mass for the neutron star. The inner disc radius is set at the co-rotation radius assuming a Keplerian disc and the outer at 9/10 of the primary's
Roche lobe.  We may estimate the mass transfer rate assuming that the low mass secondary continues to fill its Roche lobe (requiring
either/both an evolved or/and an X-ray irradiation-bloated state), and that the process
is solely driven by angular momentum loss due to gravitational radiation (GR; \pcite{kraf62}): \[\dot{M}_{GR}\simeq0.95\times10^{-11}\msun{\rm yr}^{-1}(M_2/0.05\msun)^2.\]
This is also consistent with the time-averaged value based on  the energy
released during outbursts (e.g. 1998 April, total
fluence$\simeq4.3\times10^{-3}\ergsqcm$;
\pcite{gilf98}) along with their \til20 month
recurrence cycle.
 The spectrum is plotted as the dotted line in the figure.  With a measured value for $a_1\sin i$
from pulse timing, the value of $M_2$ we choose immediately constrains the inclination angle $i$.  For a 0.05$\msun$ secondary, we find a moderately
high inclination $i=60\degree$ (50\% {\it a priori} probability of detection), and the model curve is consistent with the lower limits on the optical fluxes, and has the correct blue
slope (although $\dot{M}$ could well be slightly higher than we used).  For any higher inclination both the GR-driven mass transfer rate decreases
(as the square) and the projected disc size (as $\cos i$), making such a system too faint compared to our observations.  On the other hand, for
more massive secondaries the inclination must be lower and hence so will the {\it a priori} probability of seeing such an inclined system, ignoring any selection effects.

Depending on the system inclination a variety of mechanisms can lead to a periodic modulation of the optical flux on the binary period.  In the
case of \targ, the lack of an X-ray eclipse requires $i<82\degree$, whilst we have seen it is more likely to be lower still, $i\simlt60\degree$. In this
range low amplitude (at most a
few percent) {\it sinusoidal} modulations are seen in many LMXBs,
which are presumed to arise due to the varying contribution from the X-ray heated face of
the secondary as its aspect changes.  Within this framework the measured amplitude of modulation
enables us to estimate the level of irradiating flux required to be $L_{irr}\sim10^{33}\ergsec$.  Admittedly, this
luminosity is a factor of ten higher than that of the two quiescent X-ray observations of \targ\, but probably still consistent, given the degree of
variability seen in other quiescent sources (e.g. Cen X-4; \pcite{camp97}).  We note that this too argues against lower inclinations, as the
amplitude will decrease along with the relative contribution of the heated face, and a greater degree of irradiation is not possible.  
Interestingly, the trend in amplitude fits
that implied by the data, decreasing towards the blue, which is a consequence of the cooler variable companion star component.

\section{Conclusions}
We have undertaken the first time-series CCD photometry of V4580 Sagittarius, the optical counterpart to \targ, in quiescence.  We find that it has dropped to $V\sim21$, five magnitudes
from the peak of the 1998 outburst, although its colours are the same within the uncertainties.  Furthermore, as during outburst it exhibits a
clear 2hr sinusoidal modulation with a \til6\% semi-amplitude, and is at a minimum when the pulsar lies behind the companion. Our $BVR$ photometry
reveals a possible trend in amplitude, increasing with wavelength.  Lastly, a comparison of its X-ray to optical flux ratios in outburst and
quiescence shows that if reprocessing of the observed X-ray flux is the dominant mechanism in outburst, this cannot apply in quiescence, as it
 is relatively optically over-bright.

An analysis of these results has led us to conclude that:
\begin{itemize}
\item Unlike other soft X-ray transients in quiescence, the intrinsic emission of the companion star is not the dominant contribution to its
optical flux (as evidenced by the lack of ellipsoidal variations).
\item If we assume that a low level of mass transfer (driven by gravitational radiation losses) continues, then the disc emission due to internal
viscous energy release is the main source of the observed optical
flux.  However, the observed X-ray luminosity requires that this mass
is {\em not} being
accreted onto the neutron star.

\item The observed sinusoidal binary modulation is most consistent
with a geometrical (i.e., aspect change) origin, wherein the face of the secondary
is irradiated by the quiescent X-ray flux, and makes a variable contribution to the optical flux.
\end{itemize}

Clearly, with the existing data we have only been able to infer many of the physical details of the system and further observational effort on
the quiescent counterpart is still needed.

\section{Acknowledgments}
We are grateful to the South African Astronomical Observatory for granting time for this project and to Fran\c{c}ois van Wyck for his support at
the telescope. Financial support of LH for this work was provided by PPARC and through NASA grant NAG5-7932.

\newpage

\newpage
\newpage

\newpage

\appendix
\section{Deriving the reprocessed optical flux from the secondary's heated face}
In deriving the optical flux due to the irradiated face of the secondary star, a spherical geometry was assumed, such that the surface need only
be divided into annuli, having an angle ($\xi$) between their normal and the radiation incident from the primary. In this way, the method used for the accretion disc can be modified for the star.  Hence, the effective temperature is given
by: 
\[T^4_{irr}=\frac{(1-\eta_\star)L_{irr}\cos\xi}{4\pi\sigma d^2}\]
where $\eta_\star$ is the irradiating flux albedo (taken as 0.3; \pcite{milg76}) and $d$ is the separation of the annulus from the source.  The spectrum is
then given by assuming each annulus emits as a blackbody at this temperature, and summing the contribution from each at a given frequency (the shadowing effect of the disc has been taken into account). However, for simplicity the
line of sight is taken as lying in the orbital plane, enabling straightforward calculation of the projected surface area of each contributing
annulus.

\end{document}